Roger T. Dean, MARCS Institute for Brain, Behaviour and Development, Western Sydney University and austraLYSIS, Sydney
Jamie Forth, Department of Computing, Goldsmiths, University of London

Corresponding Author: Roger Dean. email: roger.dean@westernsydney.edu.au
ORCID: RTD, 0000_0002_8859_8902

Towards a *Deep Improviser:* a prototype deep learning post-tonal free music generator

Abstract:
Two modest-sized symbolic corpora of post-tonal and post-metric keyboard music have been constructed, one algorithmic, the other improvised. Deep learning models of each have been trained and largely optimised. Our purpose is to obtain a model with sufficient generalisation capacity that in response to a small quantity of separate fresh input seed material, it can generate outputs that are distinctive, rather than recreative of the learned corpora or the seed material. This objective has been first assessed statistically, and as judged by k-sample Anderson-Darling and Cramer tests, has been achieved. Music has been generated using the approach, and informal judgements place it roughly on a par with algorithmic and composed music in related forms. Future work will aim to enhance the model such that it can be evaluated in relation to expression, meaning and utility in real-time performance.

1. Introduction

Could a deep learning model of music function as a free-improvising partner? Free improvisation is commonly mainly post-tonal (that is, most of the time lacks emphasis on pitch structures that are hierarchical) [1,2], and often post-metrical (that is, mostly lacks hierarchical repetitive rhythmic structures) [3-5]. In other words, it is very different from common-practice Western music, or pop and rock. An experienced free improviser in any artistic form is usually capable of responding to highly diverse, often unanticipated, inputs in ways which are potentially equally diverse and sometimes unfamiliar even to the improviser [4,6] and experimental work has revealed something of the decision making involved [7-9] reviewed [10,11]. In contrast most algorithmic (computational) music generation systems have their own fixed heuristics, though sometimes responding to external events by using machine listening to transform the outputs of those otherwise unchanged heuristics [12-16]. It seems that in principle a deep learned model based on appropriate corpora might learn a diverse enough set of statistical associations that it could be generatively seeded and sampled so as to function like a free improviser. This paper demonstrates a first step towards such a system.

Previous deep learning music generation systems have mainly focused on generation of common practice instrumental music (using symbolic representations): see reviews [17-20]. A more recent emphasis is on audio generation (using digitised wave form representations) [21,22]. Considering the case of music with symbolic representation (and thus potentially conventional musical notation), the highly successful FolkRNN [23] produces music closely akin to Irish Folk music, with clear tonal and metrical features very much in common with it. Occasional audible perturbations to those features occur. Performance RNN is a recent output of the Google Magenta project [24]. It uses a quantisation of rhythm into 10msec units, in which durations up to 1sec are used, so as to create 'expressive timing' and dynamics. Pitch in the illustrative sound excerpts (as in the corpus of piano music used) are largely tonal, and the authors themselves describe the outputs somewhat dismissively as 'noodling'. One feature commonly lacking in deep learning generators is overt hierarchical structure, such as repetitions over a long time scale. In the work on audio generation from audio inputs, SampleRNN [21] introduces multiple levels of temporal hierarchy, but over less than a second. However, the principle can clearly be extended.

Given our intent to progressively assemble a *Deep Improviser,* a machine capable of free improvisation particularly in conjunction with human partners, we started with the target of multi-hand keyboard music, and considered carefully the symbolic representation appropriate for our purpose. The resultant representation is detailed in the next section.

The paper proceeds as follows. In section 2, we describe some of the long term purposes of the work, and how these lead to our adopted form of symbolic representation of keyboard music. We also summarise the key criteria we aimed to fulfil in our initial prototype system. Section 3 describes the generation of musical corpora for training the models, which in turn is described in Section 4. The generative step is discussed in Section 5, together with statistical data on the nature of the corpora, inputs and outputs. Section 6 provides discussion, conclusions and some pointers to future work.

2. Specific purposes and the musical representation

To accommodate post-tonal music, our intended model needs to permit any chord pitch combinations (playable by a single human or not!), and we decided that up to 10 pitches(notes) could be allowed in a 'chord' (while a motivic or 'melody' note would be a single pitch), based on practical experience in keyboard improvisation. Some of the relevant music involves multiple algorithmic parts, or multiple keyboard players at one or two pianos, and hence we refer to our target as 'multi-hand' keyboard music, noting that with four hands most (but not all) likely combinations of pitches, even if widespread, can be performed. To accommodate post-metrical features, our model also needs to permit continuous variation in event duration (be it chord or note), and inter-onset interval (ioi) between events (time of onset does not so clearly represent this feature, but rather requires differencing to generate it). We allowed these to be continuous variables, with an upper limit of 20 seconds (after which the sound of almost any piano note has died away). We chose further to represent pitch and key velocity (dictating acoustic intensity of the sounded note/chord) by continuous values bounded respectively by 0-120 (standard MIDI values, where the note middle C, normally termed C4, is 60), and 0-127, so that in the future these two could also be performed as continuous variables. This would allow continuously morphing microtuning, and could avoid or embrace the concept of a dominant tuning system. Note that, for example, post-tonal music does not necessarily always eschew tonality, just as 12-tone serialism, pioneered a century ago, often strives for atonality [25], but not always, and in any case does not always achieve it [1,2]. Analogous flexibilities apply to timing and tuning issues. Our system should be able to accommodate these extremes.

Thus each note or chord event is defined in the same way by a vector of 13 numbers: numbers 1-10 are pitches, so that a melody note just occupies location 1, and 2-10 are occupied by a preset value of -1, suitable for the following regression. An event comprising a chord of 5 notes would have 5 locations with pitch values and 5 locations still set at -1. We accept input pitch values from MIDI 0-120 as noted (the full range provided by some of the algorithms used in our algorithmic corpus), but only allow pitch outputs from 12-113 (many pitch values below 12 or above 113 are audible but essentially non-discriminable from each other). Input vector location 11 is occupied by a single velocity value (MIDI range 0-127), but outputs are constrained to 20-127, since values below 20 are again usually indiscriminable and mostly inaudible. Vector locations 12 and 13 in the representation provide a continuous value (quantised to integer msec in the inputs) for event duration (note or chord), and similarly a value for inter-onset interval (ioi), the delay time until the next event. Note durations may be longer than ioi, so notes may overlap each other; but our representation currently involves a simplification, in that often notes of a given chord event have different durations that we do not include. We plan to introduce this into our system later. Because the musical material being learned has time series autoregressive properties (see detail in Section 4), we included 10 lags of the input series 13-component vector as the basis for outputting a prediction of the next 13-component vector.

We want the outputs of our system to be manipulable, eventually in real time. Here, as a first step, we pursue the possibility of perturbing the outputs by means of seeding with new material (which in principle could be generated live while the model proceeds). In our preceding work on adventurous text generation (Dean and Smith, 2017, revision submitted) we also used sampling temperature (dictating the entropy of the sampling distribution from which a prediction is chosen) to enhance diversity of output, and showed the success of this. In the present work on music generation, we have delayed this manipulation for future implementation.

3. Creating multi-hand keyboard corpora

We constructed two keyboard corpora for the training of our initial models, since we can find no prior symbolic corpus with the objectives and features we required (post-tonal and post-metrical). First, our 'Algorithmic Corpus' was made by concatenating 13 runs of 6 different compositional algorithms developed in previous work by author RTD (an internationally active composer/improviser). Several of these algorithmic pieces were multi-strand in nature, that is, have multiple simultaneous melodic strands as in chamber and orchestral music, as well as chords vs melody notes. Most of the algorithms are interactive (so called live algorithms). RTD also performed the 9 keyboard improvisations specifically to formed the second, 'Improvised Corpus', using a Yamaha CP300 weighted touch sensitive keyboard at Queen Mary University of London (20170905). An excerpt of one piece is provided within the supplementary audio material, realised using the Pianoteq physical synthesis piano (which is particularly suitable because it can be used with any tuning system, not only the conventional Western tunings we use here). Our previous work showed the utility of synthetic serial music corpora in understanding the information content of such music, and in modelling its pitch features [2].

Table 1 shows the basic features of the constructed corpora; and also of the separate improvisation (recorded 2016) which was used as source of seed sequences for generation from the deep learned models. The table illustrates the distinctiveness of the three different materials in every respect bar the overlap of PCA component variances between the algorithmic corpus and the improvised seed piece. Further comparisons between the Improvised corpus and the Improvised seed piece are illustrated later in Figure 1.

Table 1. Descriptive statistics of the two constructed corpora and the seed improvisation (to be used with the models for generative purposes)

| Material | Number of events | Total number of notes | Mean notes per event | Ratio between the number of chords and the total number of events | PCA components 1 and 2, % variance explained respectively | |
|---|---|---|---|---|---|---|
| Algorithmic Corpus | 16484 | 66892 | 4.05 | 0.65 | 81.4 | 18.5 |
| Improvised Corpus | 13466 | 34397 | 2.56 | 0.56 | 70.5 | 28.7 |
| Improvised seed piece | 214 | 1001 | 4.68 | 0.72 | 81.0 | 18.7 |

Table 1 Legend. Very few chord pitch combinations are repeated at all; whereas many the pitches of many (single) melody notes are. Note that because of the occurrence of some uniform columns of -1 values (unoccupied pitch components), the PCA was done on unscaled values, for illustrative purposes only.

4. Developing a Deep Learning model

As noted above, we aimed for a model which could be perturbed (that is, sensitive to external input seeds), and generate outputs distinct in statistical nature from its learned corpus and from the seed material. So we took the avoidance of overfitting very seriously, to seek such flexibility in the generation phase. Especially given the limited size of our corpora, we considered that a model that at least beats common-sense predictions (see next paragraph for detail), but is not necessarily the most precise feasible, might have internalised statistical associations that could provide the basis for flexible generation.

As is well known, most musical time series are highly autoregressive, that is, each event is substantially predicted by a series of immediately preceding events [26-28]. Indeed, the normal common-sense prediction for time series which we adopted (often called the naïve model) is that the next event is similar to the last and so guessed as being the same [29]. We undertook some simple time series analyses on the corpora each taken as a single whole, to pre-establish that each of our features, pitch (p), velocity (v), duration (d) and inter-onset interval (ioi) are highly autoregressive, with lags of around 10 previous events being significant predictors of the next. Consequently, our representation provided inputs to the deep learning model that were sequences of 10 events, each represented by the vector of 13 values described above. These inputs were used successively to predict the next event, and the model was trained by comparing prediction with actual.

Dilated convolutional neural nets (CNN) have recently emerged as powerful models of sequence structure (see for extended review and practical tutorial on deep learning using Keras : [30]).  Given our modest recent success in using these for poetic text generation, we first considered stacked CNN alone for the Deep Improviser. Because of the apparently greater capacity of recurrent neural nets (RNN) and the LSTM (long short-term memory) nodes, we then considered RNNs receiving outputs from an initial CNN layer. We optimised models on the algorithmic corpus, and then solely tuned and fine-tuned (varying the learning rate) these for the improvised corpus, since these results were adequate for our purpose, and we were not determined as yet on utterly optimising all models.

We largely overcame overfitting by a combination of stringent dropout (parameter 0.5) at each layer, including both dropout and recurrent dropout in LSTM layers, together with L2 kernel-regularization at each layer (parameter 0.01). Overfitting was judged by continuous monitoring of loss (mean squared error) in the training set, and also in a withheld (unlearnt) validation set. These both reached their minima at a similar point, and hence the use of a second unlearnt test set was unnecessary (because the danger of progressive adaptation of the model not only to the training but also the validation set was avoided). Furthermore, given the limited size of the corpora our concern was more for the ability to generate, than of absolutely maximal learning. There was no shuffling, because these are autocorrelated time series (shuffling on the basis of pre-estimated phrase structure may be of future interest).  We used robust scaling (to address the asymmetric distributions of temporal values (shown later in Figure 1)), and undertook some hyperparameter tuning (notably to reach the minimal size nets effective for our corpora), together with limited parameter tuning using sequenced learning rates. Modeling was done with Keras and Theano. Bidirectional RNN were ineffective (as might be expected, bar the occurrence of significant retrogradation, where for example a pitch sequence may occur both forwards and backwards). Pitch augmentation (by transposing the materials to every possible relative position within an octave range) has been found effective in enhancing modelling with tonal music [31], but was not so here.

Table 2 summarises the performance of the resultant pair of models (one CNN alone, one CNN followed by RNN) as applied to both the algorithmic and improvised corpora (different weights in each case, but identical model form). The data show consistently favourable comparisons with estimates of common-sense predictions, and the capacity of the model to deal reasonably with previously unseen data (generalisability). Compared to the CNN only model, The CNN-RNN gives an improved overall rmse (root mean squared error) in the case of the Algorithmic corpus, but not with the Improvised corpus (further hyper-parameter optimisation for this case can be attempted). The data also show that the temporal features contribute a large part of the rmse, given their large values, while the rmse attached to pitch 1 is respectably small. Given pitch ranges of 0-120 in the input, it can be seen that the rmse of modelled pitch 1 in the several models (rmse range 11-19 expressed in the pitch units) constitutes a maximum error of about 16%. Especially in post-tonal music (where contour, that is whether the pitch sequence goes up or down at a given point, is likely far more important than precise pitch number) this seems to us quite usable. Note that the purpose of adding the RNN (with only a c.50% increase in model parameters) is not solely to enhance the model precision, but also in the hope of enhancing model 'memory' (autoregressive and cross-parameter temporal relationships), such that it might predict longer sequences. We return to this issue in the next section.

Table 2. Performance of the selected Deep Learning Models on the withheld validation set

| Corpus/Model | RMSE | | | | |
| --- | --- | --- | --- | --- | --- |
| | Overall (vector of 13 values) | IOI | p1 | Validation loss | Number of parameters |
| Algorithmic Corpus: | | | | | |
| Naïve model | 562.22 | 1004.23 | 24.55 | N.A. | N.A. |
| CNN only | 397.04 | 415.66 | 15.49 | 95.25 | 7565 |
| CNN/RNN | 180.36 | 488.57 | 11.54 | 96.44 | 10445 |
| | | | | | |
| Improvised Corpus: | | | | | |
| Naïve model | 254.18 | 477.91 | 19.76 | N.A. | N.A. |
| CNN only | 185.56 | 417.72 | 16.01 | 83.32 | 7565 |
| CNN/RNN | 196.30 | 450.24 | 19.20 | 80.90 | 10445 |

Table 2 Legend. The naïve model, as indicated in the text, was one in which the next event is predicted to be the same as the present one. The deep learning models were selected and optimised for hyperparameters and tuning and fine tuning on the Algorithmic corpus; then adapted by tuning and fine tuning to the Improvised corpus. CNN, 64 filters, kernel of 4, dilation 8. CNN/RNN: CNN 32 filters, kernel of 4, dilation 8, RNN LSTM 32. N.A.,not applicable.

5. Triggering and characterising distinctive outputs

Just as our models were trained on inputs of 10 events (each represented by a vector of 13 values) to predict the next event (in the same vector form), we planned to trigger outputs by using varied sequences of such vectors from an external (previously unseen) single improvised keyboard performance (1000 notes). During generation each prediction is added to the end of the seed, and the first seed member removed, so that after 10 predictions the new sequence constituting the next seed is composed entirely of model predictions. Thus one important question is how frequently should one re-seed with a sequence from the external seed; which in turn brings up the question whether the models have sufficient memory to continue generating fresh sequences ad infinitum, or whether they gradually regress to a fixed value set after a certain number of predictions: that is, gradually converge on a prediction which remains thereafter constant. We expected this regression might occur in both models, particularly the CNN only, both because of the limited capacity of CNN for temporally ordered sequences in comparison with LSTM (or GRU) units, and also the relatively small size of the learned corpora. The results of assessing this question were that both CNN and CNN/RNN models defined above regressed to a static value within 60 predictions, after seeding once. This requires that both the CNN and CNN/RNN models are reseeded very regularly for generative applications. While certain model modifications, such as the introduction of residual connections (re-injection of earlier weights) may delay this regression to the mean, it is possible that larger corpora are needed to overcome this (and in turn, they may require larger deep learning nets). Thus for assessments below both CNN and CNN/RNN models were reseeded every 10 events with a randomly chosen (thus normally new) subsequence from the seed (reseeding every 20 events was also functional). 1000 events were generated.

     All outputs were floating numbers which were approximated to the nearest integer. Rare predicted pitch and velocity values outside the ranges MIDI numbers 12-113 and 20-127 respectively, and slightly more frequent predicted duration or ioi values less than 0 msec or more than 15000 msec were rejected and the corresponding note removed. Rejection was chosen rather than rounding into range, so that these values did not distort the distributions of the respective parameters, which were to be analysed further in characterising the outputs. Based on detailed prior empirical data on improvised keyboard performance [32,7] notes that occurred within 35 msec of an initial note were grouped together as a chord.

     As we described previously (Dean and Smith, revision submitted 2017) analysis of word outputs during text generation can be done using word embeddings (vectorial representations of statistical word relations: reviewed [30]) or stylometry, based on relative word or ngram frequencies, and we used the R 'stylo' package to successfully distinguish word output distributions in our generative poetry project. While chord2vec [31] and related modifications of word2vec [33] are useful in adopting a similar approach to tonal and metrical music, they are not applicable here because specific chord voicings rarely recur (though individual notes of course do), and the whole vector of p1-10,v,d,ioi essentially never recurs exactly partly because of the continuous parameters involved in d and ioi. Thus an alternative approach to assessing whether outputs are distinctive or merely recreative has been adopted, in which we undertake univariate and multivariate testing of the question: what is the probability that the distribution of pitch (or velocity, duration, ioi) values observed in one case (e.g. the Algorithmic corpus) and that observed in another (e.g. the generated output from the Algorithmic corpus when seeded with the external improvised sequences) both arise from a parent distribution (which remains unspecified in nature). For this purpose (and for playback) we reorganised the pitches of each chord into MIDI sequences of individual notes (for this assessment we are not considering time series sequential relationships). The Anderson-Darling test can determine this value, and Table 3 shows as an example the relevant results from the simpler CNN-only model based on the Algorithmic and on the Improvised corpora, confirming that seeding with the external sequences from an improvised

piece (not included within the Improvised corpus) is effective in driving generation which is not simply distributed the same way as either the learned corpus or the seed. The multivariate Cramer test, assessing whether two distributions are distinct, supports this conclusion where undertaken, considering all the parameters p1-10, v, d, ioi simultaneously in their chord/note event representation. Because the R Cramer algorithm generates very large numbers (which can outstrip the range permissible with the 32 bit number representations that R normally expects) it was necessary to undertake the test with subsets of the data. For example, it gave $p = 0$ for comparing 3000 sequential notes from the Improvised Corpus, and from the deep learned CNN model of that corpus, seeded by the external improvised piece.

Table 3. Output Distinctiveness: Univariate testing for possible statistical common origins of corpora, seed, and generated outputs, based on note sequences.

| Generator Model | Comparison Distribution Origin | Anderson-Darling statistic T.AD, and (probability of origin from a common distribution reported by R) for the specified distribution parameter | | | |
|---|---|---|---|---|---|
| | | pitch distribution | key velocity distribution | note duration | note inter-onset interval |
| Algorithmic Corpus | Generated output | 734.2 (<0.001) | 394.7 (<0.001) | 615.8 (<0.001) | 1370 (0) |
| | Input seed | 1090 (0) | 1123 (0) | 94.72 (<0.001) | 138.3 (<0.001) |
| Improvised Corpus | Generated output | 135.4 (<0.001) | 2670 (0) | 694.3 (<0.001) | 2210 (0) |
| | Input seed | 909.1 (0) | 666.4 (<0.001) | 146.6 (<0.001) | 203.2 (<0.001) |

Table 3 Legend. The CNN-only models of the Algorithmic or the Improvised Corpus was seeded with sequences of the external improvised piece (not included in the Improvised Corpus). The properties of the corpora and the seed are summarised in Table 1. The Anderson-Darling kSample test was done in R with package kSamples. Its statistic, T.AD is (AD criterion – mean)/sd and there are separate versions for discrete and continuous distributions (quoted accordingly above). The statistic then provides a probability that the two distributions considered could come from a shared parental distribution (whose nature is not determined). These univariate measures were based on notes (and not chords, for reasons discussed in the text). The corpora and the seed were also mutually distinct judged by this test. p, pitch (encompassing all of p1-p10 sounded pitches, and disregarding the unsounded values), v attack velocity, d note duration, ioi interonset interval.

Figure 1 illustrates some of the distributions that are included in Table 3, specifically those from the seeded Improvised Corpus modelled by the CNN-only, and shows that its implications are visually plausible. Separate Anderson-Darling analyses (not shown) of the outputs from the CNN/RNN model of the algorithmic corpus (and the parallel analyses with the CNN-RNN Improvised corpus model) support a general conclusion: that our approach permits the generation of sequences statistically distinct from either the learned corpus or the input seed distribution.

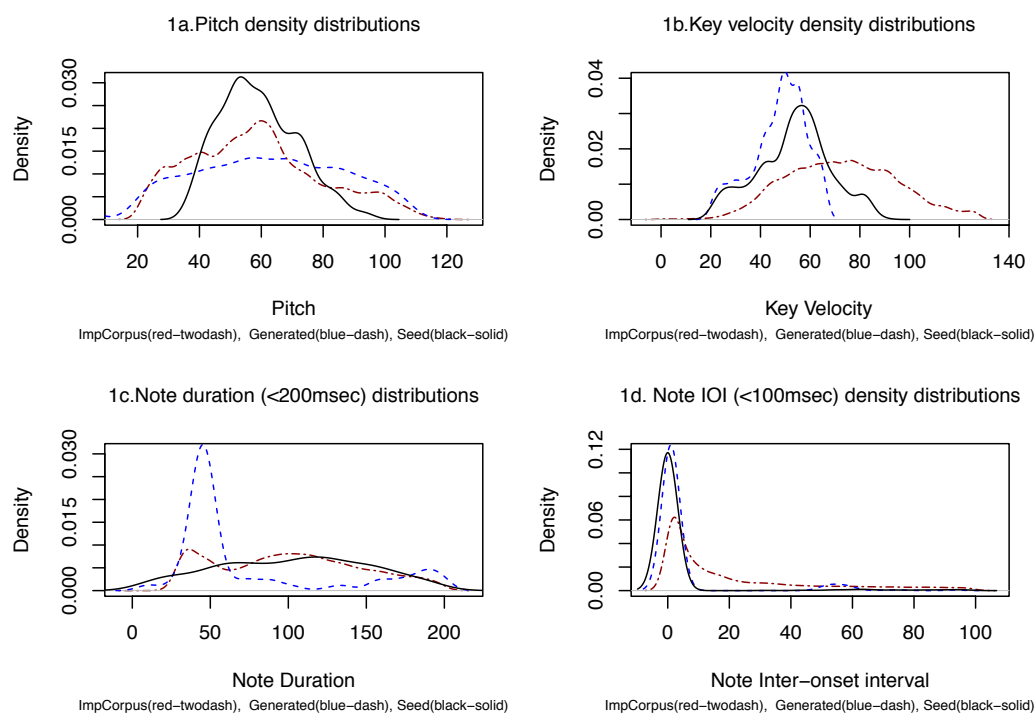

Figure 1. Density distributions of Pitch (a), key velocity (b), note duration(c) and note IOI (inter-onset intervals: d) for the Improvised Corpus, the Seed, and the Generated output (1000 events). The descriptive characteristics of the Corpus and the Seed are summarised in Table 1. The distinctiveness of the generated material is revealed as partly due to its broad pitch distribution (1a), and partly to its emphasis on lower key velocities than either the seed or corpus(1b). It has a dominant mass of durations around 50msec (1c), in part because of the simplification introduced in the model representation whereby the notes of a chord all have the same duration. This effect is supported by its relative preponderance of chords over melody notes, resulting in a large mass around 0 msec note inter-onset interval, shared with the seed but not the corpus (1d). Note that the graphed note duration distributions are truncated at 200ms, and the inter-onset interval distributions at 100msec to make the distinctions as clear as possible (no distinctions are visible in the much longer time values, and for example the IOI distributions range up to almost 20000 ms).

Going from statistical distinction to substantive human evaluation of computational artistic generativity is a hugely difficult task [34-37] as for that matter is evaluation of (manually) composed work; and there is also an argument that computational creativity (to which this paper potentially contributes) should be assessed in relation to its own specified objectives, partly or even solely by an internal mechanism [38]. In a previous paper on generativity with time series models [39] we elaborated on possible approaches which could minimise the psychological 'demand' commonly imposed in human listening tests (for example, avoiding reference to computational vs compositional origins and hence the biases these commonly elicit). In addition, we pointed to the complexity of evaluating an output which occurs simultaneously with a live improvisation or an enunciation of a pre-composed element. While we intend to grapple fully with this in due course, for the time being we allowed ourselves a preliminary informal test: 21 researchers listened to 3 unidentified items of post-tonal and post-metrical keyboard music in a group setting. The items were 75-100 seconds in length, and presented as 'multi-hand keyboard music' (and so it was pointed out that it was not necessary to assess whether what was heard was feasible for a single human to play). Listeners were informed that one piece was composed manually, one was composed algorithmically, and one was generated by a deep learning model with seeding. The first item was an extract of Morton Feldman's *Piano Four Hands* (1957-8, from the Etcetera KTC2015 CD performed by Roger Woodward), the second an extract of a Deep Improviser product, and the third taken from an algorithmic piece included in the corpus it had learned (the latter two extracts are provided as supplementary material, while the first cannot be published here for reasons of copyright, but is available). After all items had been heard, votes were called first for preferred item, being respectively 3:3:15, and then for which piece was the deep learned product (11:10:0). There were only a few people amongst this group who had prior exposure to music of this kind, but in a separate group of three friends, all familiar with such work, when given the same test and scenario, the scores were respectively 1:1:1, and 1:0:2. Thus the Deep Improviser was not readily identified, and the algorithmic piece was preferred amongst the three, but the Deep Improviser was competitive with the human composer. We view this at least as a modicum of support for the utility of our approach (RTD is a great admirer and was an acquaintance of the composer).

6. Discussion, Conclusions and Future Work

Overall, our prototype *Deep Improviser* shows initial signs of success: it can generate outputs distinct from its learned corpus or input seeds that nevertheless have commonality with them. Clearly this distinction can likely be dramatically enhanced by control of sampling temperature. There are numerous limitations to the model so far, of which perhaps the most obvious are the relatively small corpora (though this is an advantage for any music maker wanting to establish their own model and corpora), and the sparseness of occupancy of the p1-10 part of the vectorial representation of an event (together with the lack of distinct duration values for individual notes of a chord). On the other hand, one-hot encoding (where for example pitches 12-113 would be represented each by a vector of 101 zeros with a 1 at the spot in the vector corresponding to the pitch) is also a sparse representation, yet has many benefits including categorical prediction, and may be useful here (it is used as the entire basis of Performance RNN encoding). Quantising durations, for example at the observed 35msec cut off between a succession of notes and a chord, may also be valuable. The fact that MIDI represents chords as a succession of notes, separated by 0 or a few msec in timing also indicates that considering we have to interconvert recorded chords and notes, there may be a case for generating chords as sequences of notes rather than as such, and modelling accordingly. This would also invite a hierarchical conditional model in which the first prediction is whether an event is a note or a chord, and the subsequent predictions then evaluate the chosen case (note or chord expressed as a rapid sequence of notes). This may ensure a wider range in the relative occurrence of chord vs. melody notes in the generated outputs: currently the chord : note ratio is commonly quite high. It would also present a pathway towards production of multiple parallel streams of events.

The time series regressions assessing relations between p, v, d, and ioi mentioned above also revealed limited mutual influences in comparison with autoregressive influences, with minor exceptions. This suggests that a multi-input multi-output (branched) deep learning model, with only certain influences permitted, may provide a more accurate model, and one in which the weighting of the loss determined on the different components of the prediction (p,v,d,ioi) might be varied according to their variability and relative importance, analogous to approaches developed with multidimensional Markov models of music such as IDyOM [40]. This is underpinned by the fact that the distributions of the temporal features are very different from those of pitch and velocity. We will consider transformations based on cumulative density functions, identification of repetitions and geometricals structures within perceptually grounded representations as possible means of reducing the complexity of those data [41,42] particular for further analyses of outputs.

In the future development of this project, we want to create and use a system operative in real time, and particularly given pre-learned models, this is already feasible. It will also be possible to fit and update models in real-time over an accumulating performed input, at the same time as generating from the current model with seeding and sampling. In our previous work we have demonstrated the utility of analytical autoregressive multivariate time series models as generators themselves, and constructed a system operative in real time [39]. The present *Deep Improviser*, while yet shallow, when comprised of CNN and RNN probably imitates some of the sequential aspects of time series models. But we also plan in the longer run to integrate *Deep Improviser* with algorithmic approaches based on information theoretic and perceptual decision-making models [43,44].

7. Electronic Supplementary Material (ESM).
Online Resource 1. An extract of generation by the CNN model of the algorithmic corpus, with seeding (ESM1-dlgen.mp3). MIDI file recorded to audio using Pianoteq.

Online Resource 2. An extract of an algorithmic piece by RTD included in the Algorithmic corpus. (ESM2-algo.mp3). MIDI file recorded to audio using Pianoteq.

8.Conflict of Interest. The authors declare they have no conflict of interest.

9.References